\documentclass[sigconf]{cidr-2023}

\usepackage{tikz}
\usepackage{amsmath}
\usepackage{url}
\usepackage{booktabs} 

\usepackage{xspace}
\usepackage{color}
\usepackage[small,compact]{titlesec}
\usepackage{tabularx}
\usepackage[labelfont=bf,font=small]{caption}
\usepackage{tcolorbox}
\usepackage[T1]{fontenc}
\usepackage{fancyvrb}
\usepackage{subcaption}     

\newcommand{\sn}{\texttt{TROD}\xspace}

\usepackage{enumitem}
\setitemize{topsep=2pt, itemsep=0.3em, leftmargin=1em}
\setenumerate{topsep=2pt, itemsep=0.3em, leftmargin=2em}

\usepackage{listings}
\definecolor{codegreen}{rgb}{0,0.6,0}
\definecolor{codeblue}{rgb}{0,0.5,1.0}
\definecolor{codegray}{rgb}{0.5,0.5,0.5}
\definecolor{codepurple}{rgb}{0.58,0,0.82}
\definecolor{backcolour}{rgb}{0.95,0.95,0.92}

\lstdefinestyle{mystyle}{
    commentstyle=\color{codeblue},
    keywordstyle=\color{magenta},
    numberstyle=\tiny\color{codegray},
    stringstyle=\color{codepurple},
    basicstyle=\ttfamily\small,
    breakatwhitespace=false,
    breaklines=true,
    captionpos=b,
    keepspaces=true,
    numbers=left,
    numbersep=5pt,
    showspaces=false,
    showstringspaces=false,
    showtabs=false,
    tabsize=2,
    xleftmargin=6mm
}

\vfuzz \hfuzz

\begin{document}

\title{Transactions Make Debugging Easy}

\author{Qian Li$^1$, Peter Kraft$^1$, Michael Cafarella$^2$, \c{C}a\u{g}atay Demiralp$^3$, Goetz Graefe$^4$, Christos Kozyrakis$^1$, \\
Michael Stonebraker$^2$, Lalith Suresh$^5$, and Matei Zaharia$^1$}
\affiliation{%
  \institution{$^1$Stanford, $^2$MIT, $^3$Sigma Computing, $^4$Google, $^5$VMware}
  \country{}
}

\renewcommand{\shortauthors}{Li et al.}
\renewcommand{\shorttitle}{Transactions Make Debugging Easy}

\begin{abstract}
We propose \sn, a novel transaction-oriented framework for debugging modern distributed web applications and online services.
Our critical insight is that if applications store all state in databases and only access state transactionally,
\sn can use lightweight always-on tracing to track the history of application state changes and data provenance, and then leverage the captured traces and transaction logs to faithfully replay or even test modified code retroactively on any past event. 
We demonstrate how \sn can simplify programming and debugging in production applications, list several research challenges and directions, and encourage the database and systems communities to drastically rethink the synergy between the way people develop and debug applications.
\end{abstract}

\maketitle

\vspace{-3mm}
\section{Introduction}
\label{sec:introduction}

In this paper, we propose \sn, a novel \textbf{\underline{Tr}ansaction-\underline{O}riented \underline{D}ebugging} framework for modern distributed web applications such as a travel reservation website or an e-commerce microservices application.
\sn targets applications that follow three design principles:

\begin{enumerate}[label=\textbf{P\arabic*.}]
    \item Store all application-shared state in databases.
    \item Access or update shared state only through ACID transactions.
    \item Produce deterministic outputs and state changes.
\end{enumerate}

We adopt these principles because they radically simplify the problem of debugging modern distributed applications.
Currently, debugging is hard because developers need to unravel the complex interactions of thousands of concurrent events~\cite{whittaker2018debugging}.
Existing distributed debugging tools are limited as they rely on developers to provide sufficient logs and traces~\cite{zhang2017pensieve}, which requires intensive manual logging or annotations.
However, if applications follow our principles, \sn can augment database transaction logging to capture a complete record of application state accesses and changes, enabling powerful features such as faithful replay of any past event.

The \sn principles are practical because they align with current trends in application design. 
For example, developers increasingly deploy applications on serverless platforms such as AWS Lambda~\cite{awslambda}. 
These serverless applications naturally follow \sn principles because they handle requests with stateless and deterministic functions and manage state using cloud databases.
In fact, we originally developed \sn to debug applications in the DBMS-oriented operating system (DBOS) project~\cite{2022dbos-progress}, which runs programs as workflows of transactional serverless functions~\cite{apiary}.
As part of DBOS, we found it was easy to build performant web and microservice applications using \sn principles.

\sn helps developers investigate and better understand bugs by faithfully replaying past events.
Faithful replay is challenging because we need to accurately reconstruct past state for applications while considering interleaving concurrent executions.
However, because \sn assumes state is centralized in databases and only accessed transactionally,
it can capture a detailed history of application state changes and events with lightweight always-on transaction tracing.
Developers can directly query traces to locate buggy executions. 
Then, during replay, \sn can re-apply logged state changes to reconstruct past application states.
Since \sn can consistently replay past events,
it transforms most Heisenbugs~\cite{gray1986computers},
bugs that happen rarely and are hard to reproduce because of complex and unpredictable interactions between concurrent events,
into easily reproducible ``Bohrbugs''.

\sn extends faithful replay to support an even more powerful debugging feature: retroactive programming.
Developers can use \sn to test their modified code on past events,
for example, to test a bug fix before pushing it into production.
Retroactive programming is challenging because \sn cannot simply re-apply the transaction log as in replay 
but must actually re-execute all concurrent events as their computations and effects might change.
Retroactive programming is possible in \sn because request executions only share state through transactions.
Therefore, \sn can identify relevant transactions and only enumerate possible re-execution orderings of those transactions to thoroughly test different possible effects of interleaving concurrent executions.

Our proposed implementation of \sn mainly focuses on correctness issues such as functionality or semantics bugs,
with an emphasis on hard-to-debug server-side concurrency issues~\cite{qiu2022deep}.
At present, we have much of \sn running in DBOS.
Preliminary results are promising, including low overhead (<15\%) always-on tracing.
We expect to present a demonstration at the conference.
\section{Debugging Frustrations}
\label{sec:background}

\begin{figure}[t]
\begin{lstlisting}[linewidth=\columnwidth,language=Python,basicstyle=\footnotesize\fontfamily{lmtt}\fontseries{m}\selectfont,numberstyle=\small,tabsize=2,aboveskip=0pt,belowskip=0pt]
def subscribeUser(userId, forum):
  # 1st transaction: Check subscription.
  if (isSubscribed(userId, forum)):
    return True
  # 2nd transaction: Insert a subscription entry.
  result = DB.insert("forum_sub", userId, forum)
  return result

def fetchSubscribers(forum):
  # Error: Duplicated values in column userId.
  result = DB.executeQuery("SELECT userId FROM forum_sub WHERE forum=forum")
  return result
\end{lstlisting}
\vspace{-2mm}
\caption{Simplified code for a concurrency bug in the database-backed online education platform Moodle (MDL-59854). The first handler, \texttt{subscribeUser}, has two transactions and manifests a race condition causing duplicated subscriptions but returns with no error.
The second handler, \texttt{fetchSubscribers}, fetches the list of subscribers and raises an error if it detects duplicates. }
\label{fig:example-code}
\end{figure}

Debugging applications that serve many concurrent requests, such as web services or microservices,
is difficult because developers need to unravel the complex interleaving of concurrent executions.
We illustrate real debugging frustrations using a concurrency bug (Figure~\ref{fig:example-code}) reported in the popular database-backed online education platform Moodle\footnote{Moodle stores shared application state in a relational SQL database and uses transactions to manage state. It supports MySQL, Postgres, Microsoft SQL Server, or Oracle.}: MDL-59854~\cite{moodle-59854}.
This is a time-of-check to time-of-use (TOCTOU) bug: a race condition exists in the \texttt{subscribeUser} handler between checking if a subscription exists (1st transaction) and inserting a new subscription into the database (2nd transaction).
However, debugging this issue is tricky because it only surfaces if two requests for the same user to subscribe to the same forum are interleaved in a certain way, and an error is only raised on a subsequent request to fetch a list of subscribers to that forum.
The developer who reported this bug commented:
``\emph{You have to be pretty fast and pretty lucky to actually reproduce this issue.}''

Locating and reproducing concurrency bugs like MDL-59854 is fundamentally challenging because these bugs are caused by specific and rare interleavings of concurrent events that are seemingly unrelated.
Conventional error messages and stack traces, as commonly used in bug reports,
are not sufficient because they only provide information on requests that fail but not on other requests that may be involved in the bug.
For example, in MDL-59854, logs and stack traces did not explain why duplicated entries appeared in the first place, and did not reference the original interleaved execution order that caused the issue.
Instead, developers had to guess which requests and functions may have inserted or updated entries to the affected database table,
and struggled to reproduce the bug because they lacked debugger support to faithfully replay past events.

Based on this and other bug reports that we examined,
we observe that to find and reproduce bugs, developers must often collect information from multiple application production logs as well as database logs.
Debugging would be easier if developers had a full view of application execution and database interactions and could replay past events.
To make this possible, we are building \sn.
\section{\sn: Transaction-Oriented Debugger}
\label{sec:vision}

\subsection{\sn Overview}
Our \sn proposal focuses on debugging correctness issues in distributed web applications and online services such as a travel reservation web service or e-commerce microservices application.
These applications typically implement their business logic in backend servers that consist of many request handlers.
When a request arrives at the backend, the application runtime invokes the corresponding request handler to serve the request.
Many applications implement a microservice architecture~\cite{laigner2021data},
so to serve a single user request, a request handler may invoke multiple other request handlers through RPCs, forming a workflow of handler invocations.
\sn assumes that applications propagate a unique ID for each request (\texttt{ReqId}) through RPCs, which is a common practice.

\sn requires applications follow three design principles (Section~\ref{sec:introduction}).
They should store all shared state and data in a transactional database (\textbf{P1}),
and request handlers should only access or update shared state through ACID transactions (\textbf{P2}).
Additionally, each request handler should be deterministic (\textbf{P3}),
where its output and state changes are determined only by the input and the database state at the time of its execution.
Request handlers can maintain local private state for individual requests, but otherwise access or persist state across requests only through databases.
\sn can work for applications using multiple data stores if their transaction logs are aligned (e.g., using a cross-data store transaction manager~\cite{cherrygarcia,faria2021towards}), but in the rest of the paper, we assume applications use a single DBMS for simplicity.

\begin{figure}[t]
	\centering
	\includegraphics[width=\linewidth]{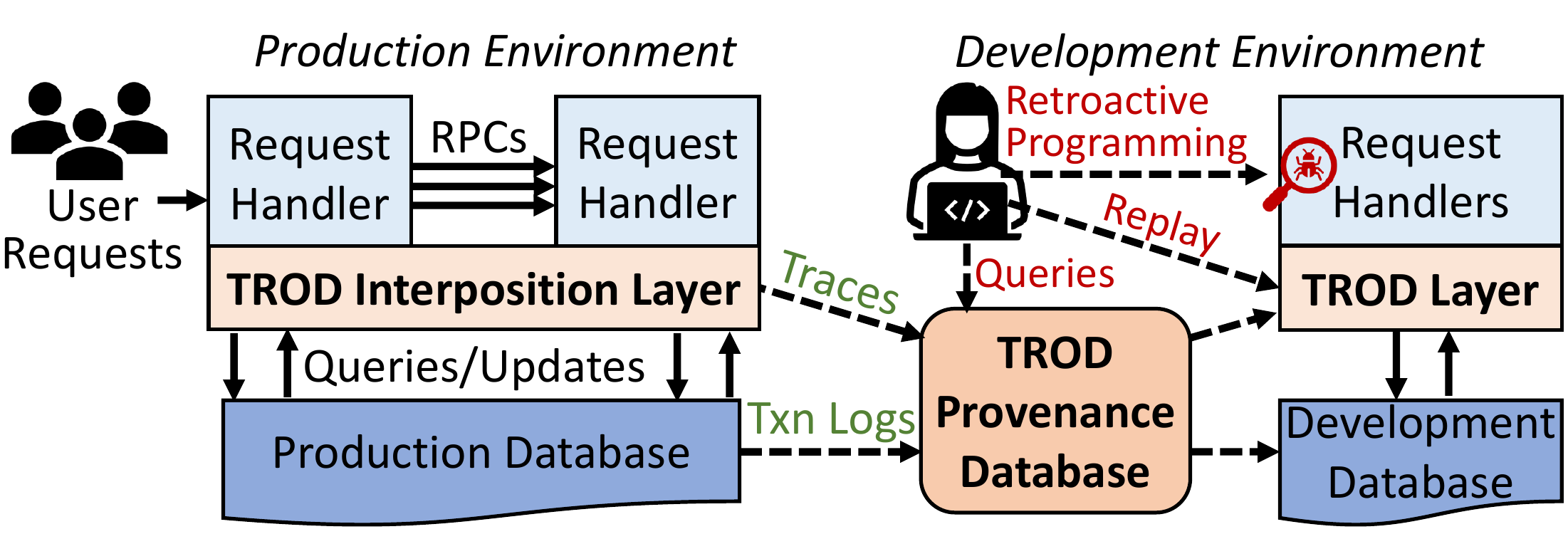}
	\caption{\sn's integration with production and development environments. \sn's tracing flows and debugging operations are marked with dashed arrows and colored text.
	}
	\label{fig:system_architecture}
\end{figure}

As shown in Figure~\ref{fig:system_architecture}, \sn has two major components:
\begin{itemize}
  \item \emph{Interposition Layer}: \sn adds a thin shim layer that hooks into the developer's application framework. It interposes on every handler and database query to trace application-database interactions, including transaction execution orders and information on which data items were read from or written to the database.
  \item \emph{Provenance Database}: \sn utilizes an analytical database to store traced data. \sn uses it to replay past events; developers can also directly query this database for debugging. 
\end{itemize}

\noindent\textbf{Simplifying Assumptions.}
We assume external service calls are idempotent, so re-executions will not generate unexpected side effects (e.g., send an email twice).
We also assume that machines are reliable and have enough hardware resources,
so resource exhaustion is out of the scope of this paper.
We leave proper handling of external services and hardware environments for future work.

Additionally, we assume for simplicity that the database provides strict serializability,
meaning transactions are serializable and serialized in commit order~\cite{linearizability}.
However, \sn can work for lower isolation levels such as snapshot isolation and read committed by leveraging prior work on \emph{transaction reenactment}~\cite{arab2014generic}, which can faithfully replay transactional histories under weak isolation levels using database audit logs and time travel capabilities.

\subsection{\sn Principles}
\label{sec:principles}

\sn design principles make debugging easier and more efficient because they exploit the synergy between application development and debugging.
If request handlers are deterministic and only access shared state transactionally through a database,
\sn can automatically intercept queries and trace data operations to capture provenance information.
Then, during replay, \sn can efficiently restore application state by re-applying the captured sequence of changes made to the database, without needing to replay handler executions.
By contrast, if an application maintains shared state outside of the database, such as in files or in-memory data structures, obtaining provenance information requires low-level code instrumentation or intensive manual annotations and can be prohibitively expensive.

\sn principles are practical because they align with two important trends in application development.
First, developers increasingly deploy applications on serverless platforms (e.g., AWS Lambda~\cite{awslambda}).
Serverless applications naturally follow \sn principles because they handle requests with stateless and deterministic functions and manage state using cloud databases.
Second, due to the growing demand for strong consistency, data stores are increasingly adding transactional guarantees while providing high performance~\cite{zhou2021foundationdb}.
For example, many NoSQL data stores now support ACID transactions (e.g., FoundationDB, MongoDB).
Therefore, it will only become easier for developers to adopt \sn principles without substantially changing their applications or data stores.

\subsection{\sn in Action}
\label{sec:in-action}
To show how \sn works, we describe how a developer could use it to debug our earlier example MDL-59584 (Section~\ref{sec:background}), a bug that caused duplicated entries in Moodle's forum subscription table.
A major challenge in solving this bug was figuring out which operation created the duplicate entries.
In \sn, this is easy: developers directly query \sn's provenance database to track which prior requests inserted the duplicated records.
\sn automatically traces essential provenance information such as request IDs, timestamps, handler names, transaction IDs,
and transaction read and write sets, storing this information in structured database tables
(we discuss details in Section~\ref{sec:automatic-tracing}).
Here is an example query to find the requests and corresponding handlers that inserted the duplicated records:
\begin{tcolorbox}[left=0pt,right=0pt,top=0pt,bottom=0pt,before skip=5pt,after skip=5pt]
	\begin{Verbatim}[fontfamily=lmtt,fontsize=\small]
SELECT Timestamp, ReqId, HandlerName
FROM Executions as E, ForumEvents as F
  ON E.TxnId = F.TxnId
WHERE F.UserId = 'U1' AND F.Forum = 'F2'
  AND F.Type = 'Insert'
ORDER BY Timestamp ASC;
	\end{Verbatim}
\end{tcolorbox}

This query returns two different request IDs with the same handler name (``subscribeUser'') and adjacent timestamps,
indicating a potential concurrency bug in the \texttt{subscribeUser} handler.
Developers can give this information to \sn to replay (Section~\ref{sec:bug-replay}) the requests that inserted the duplicated subscription.
During replay, \sn uses its captured provenance data to provide replayed transactions with the correct database state/context by interposing between handlers and the database.
Developers can then use low-level debugger tools like GDB to add breakpoints in the handler code and step through the replayed request execution.

After identifying and fixing the bug, developers can test their bugfix patch with \sn's retroactive programming feature (Section~\ref{sec:retroactive-programming}).
For instance, one developer suggested~\cite{moodle-59854} that \texttt{isSubscribed} and \texttt{DB.insert} should be wrapped in one transaction.
Thus, they can modify \texttt{subscribeUser} accordingly and use \sn to re-execute request handlers to test the original two conflicting subscription requests over a past snapshot,
and observe if the patch actually fixes the duplication issue.

\subsection{Always-On Tracing and Declarative Debugging}
\label{sec:automatic-tracing}
To collect information needed for debugging,
\sn's interposition layer automatically interposes on each request handler executed during applications' normal execution in production environments, collecting request IDs, handler names, and execution timestamps.
\sn also interposes on every operation in the application database to record which request handler invokes which transactions and what data is read from or written to the database.
Thus, \sn can trace the flow of data through handler executions, recording which transactions are part of which handler executions and which data items they read or write.
\sn organizes this captured \emph{provenance} data into structured tables in an easily accessible and queryable provenance database.
Our prototype of this always-on tracing achieves low runtime overhead in DBOS (Section~\ref{sec:prototype}), indicating it is practical.

We illustrate \sn's provenance logs using MDL-59854.
\sn records transaction executions in the \texttt{Invocations} table (Table~\ref{tab:exec_logs}),
which contains both transaction information (e.g., transaction IDs and timestamps) and request handler metadata (e.g., the handler and function names that initiated the transaction).
Records for a request share the same \texttt{ReqId}, so developers can easily observe the handler and transaction execution order within and across requests.
Developers can query this table to inspect execution histories, such as finding all transactions involved in serving a request.

\begin{table}[h!]
    \small
    \setlength{\tabcolsep}{3pt}
    \begin{tabular}{@{}c c c c c@{}}
     \toprule
     \textbf{TxnId} & \textbf{Timestamp} & \textbf{HandlerName} & \textbf{ReqId} & \textbf{Metadata} \\ 
     \midrule
     TXN1 & TS1 & subscribeUser & R1 & func:isSubscribed \\ 
     TXN2 & TS2 & subscribeUser & R2 & func:isSubscribed \\
     TXN3 & TS3 & subscribeUser & R2 & func:DB.insert \\
     TXN4 & TS4 & subscribeUser & R1 & func:DB.insert \\
     TXN9 & TS9 & fetchSubscribers & R3 & func:DB.executeQuery \\
     \bottomrule
    \end{tabular}
    \caption{An illustrative transaction execution log.}
\label{tab:exec_logs}
\vspace{-6mm}
\end{table}

For each application table, \sn tracks data provenance such as what data items are read or written by each transaction.
For data writes, \sn leverages the change data capture feature provided by most of the databases.
For data reads, \sn can use existing database provenance techniques to rewrite queries~\cite{arab2014generic} and record read set automatically.
For example, assume the provenance table name for forum subscription is \texttt{ForumEvents} (Table~\ref{tab:dataop_logs}).
\begin{table}[h!]
    \small
    \begin{tabular}{@{}c c c c c@{}}
     \toprule
     \textbf{TxnId} & \textbf{Type} & \textbf{Query} & \textbf{UserId} & \textbf{Forum} \\ 
     \midrule
     TXN1 & Read & Check if (U1, F2) exists & null & null \\ 
     TXN2 & Read & Check if (U1, F2) exists & null & null \\
     TXN3 & Insert & Insert (U1, F2) & U1 & F2 \\
     TXN4 & Insert & Insert (U1, F2) & U1 & F2 \\
     TXN9 & Read & Select UserId for F2 & U1 & F2 \\
     TXN9 & Read & Select UserId for F2 & U1 & F2 \\
     \bottomrule
    \end{tabular}
    \caption{An illustrative data operations log.}
\label{tab:dataop_logs}
\vspace{-6mm}
\end{table}

Developers can use a declarative language like SQL to directly query \sn's captured data for debugging,
which simplifies locating the root causes of bugs.
For example, if we run the query shown in Section~\ref{sec:in-action} over Tables~\ref{tab:exec_logs} and \ref{tab:dataop_logs},
\sn correctly returns information identifying the buggy request executions: (TS3, R2, subscribeUser) and (TS4, R1, subscribeUser).

\subsection{Bug Replay}
\label{sec:bug-replay}

Using \sn, developers can easily replay past requests in a development environment.
Our key insight is that if handlers are deterministic and access shared state only transactionally, \sn can faithfully replay a past execution by re-executing its code normally but restoring the database before each transaction to a state equivalent to what the transaction originally saw.
To make this possible, \sn adds a breakpoint before the beginning of each transaction.
Within a breakpoint, \sn queries the provenance database to find what state changes the upcoming transaction depends on, and then applies those changes to the development database.
Therefore, the re-executed transaction can see the same state and return the same result as the original execution.
\sn then repeats the above steps at each breakpoint until all handlers and transactions finish.

\sn's replay is practical and easy to use because it can faithfully replay production events in a smaller development environment and is compatible with other debugging tools.
\sn can efficiently replay request executions without needing to restore the entire production database
because it can use captured provenance information to only restore those data items used in replayed transactions.
During replay, \sn lets developers use familiar low-level debugging tools to inspect executions in detail.
For example, developers can use GDB to add breakpoints and single-step through a replayed handler execution,
or use built-in database monitoring tools to observe detailed query plans.
\sn augments these debugging tools with new capabilities.
For example, if a request spans multiple transactions, \sn makes it easy for developers to query which concurrent executions may have updated the database between transactions, making many concurrency bugs easier to understand.

To debug the Moodle issue (MDL-59854) with \sn, developers can replay the request R1 that caused the duplication. 
\sn replays the original execution order of R1, shown in Figure~\ref{fig:retro_debugging} (top).
First, \sn restores the database to a snapshot before R1 execution and replays R1's first transaction \texttt{isSubscribed}, which finds that no subscription exists.
Then, \sn injects the relevant database change made by request R2, which inserts a subscription entry (U1, F2).
Next, \sn re-issues R1's second transaction to the database and in turn inserts the duplicated entry.
During replay, developers can see that the database was modified by R2 between the executions of R1's two transactions, indicating the cause of the bug.
Therefore, \sn provides detailed information for developers so they can better understand how duplication occurs: interleaving executions of transactions from two requests.

\subsection{Retroactive Programming}
\label{sec:retroactive-programming}

\sn not only lets developers faithfully replay past executions, but also lets them \emph{retroactively program} applications, modifying code and testing it on past events.
Retroactive programming is important because it helps developers verify that a proposed bugfix patch actually fixes the bug and does not introduce new bugs.

Retroactive programming is challenging because to fix bugs, developers typically have to change application logic.
For instance, developers may split a long transaction into multiple shorter ones or combine several small transactions into a big one.
Therefore, \sn cannot re-apply the transaction log as in bug replay, but must actually re-execute all relevant concurrent requests.
It is not obvious in which order to execute those concurrent requests, especially if their logic and transaction boundaries have been changed.
Naively, there are a prohibitively large number of possible ways to interleave instructions among concurrent executions.
However, since \sn requires handlers only share state through transactions,
\sn can identify relevant transactions (e.g., transactions that access the same table) and enumerate possible re-execution orderings among them to thoroughly test different possible effects of interleaving concurrent executions.

\begin{figure}[t]
	\centering
	\includegraphics[width=\linewidth]{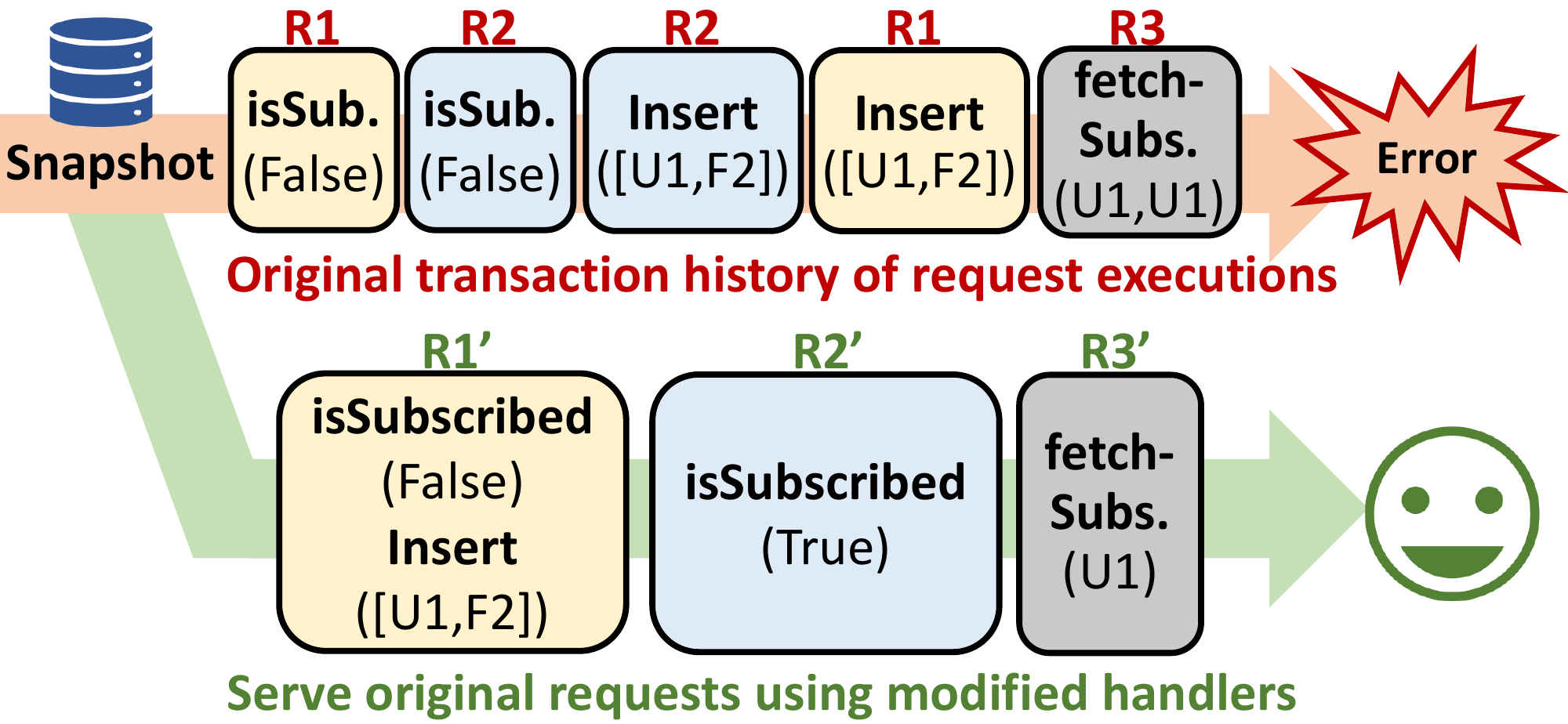}
	\caption{Top: Original transaction log. Bottom: Retroactive programming log where \sn executes the modified code to serve original user requests on a past database snapshot.
  We color boxes handling the same request with the same color, and put the return values in parentheses under each function name.
	}
	\label{fig:retro_debugging}
\end{figure}

To test the bugfix patch for the Moodle issue (MDL-59854), developers can use \sn to re-execute original requests R1, R2, and R3 using the modified code where the \texttt{subscribeUser} handler now wraps \texttt{isSubscribed} and \texttt{DB.insert} in one transaction.
First, \sn restores the development database to a snapshot right before R1 execution.
Then, \sn branches off from the original history to execute and trace new executions.
\sn tests two possible re-execution orders for the concurrent requests (R1' first or R2' first), and we show one of them in Figure~\ref{fig:retro_debugging} (bottom).
After both R1' and R2' finish, \sn executes R3'.
Developers can observe that the third request which fetches subscribed users no longer raises any errors.

\subsection{\sn Prototype}
\label{sec:prototype}
We are currently prototyping \sn in DBOS.
As introduced in our recent paper~\cite{apiary}, we implement always-on tracing using a high-performance in-memory buffer and find that on popular microservices benchmarks, the overall tracing overhead is <100$\mu$s per request. This causes a relative overhead of <15\% when using the in-memory database VoltDB and negligible overhead when using the on-disk database Postgres.
We also run declarative debugging queries over billions of events and get results in <5 seconds, which is promising for interactive debugging.
We are currently implementing bug replay and retroactive programming, and expect to give a demonstration at the conference.
\section{Case Studies}
\label{sec:case-study}

In this section, we examine two categories of bugs:
1) concurrency bugs, which can be debugged through \sn bug replay and retroactive programming, and 
2) security bugs, which can be debugged using \sn tracing and declarative debugging.

\subsection{Concurrency Bugs}
To show how \sn can simplify debugging,
we examine concurrency bugs discussed in prior work on non-reproducible bugs~\cite{erfani2014works} and server-side request races~\cite{qiu2022deep}.
In particular, we look at bug reports from two popular database-backed web applications: MediaWiki (MW) and Moodle (MDL).
MediaWiki is an open-source platform underlying Wikipedia and related sites, while Moodle is an online education web platform; both have millions of users.
Both applications use a relational SQL database, such as MySQL and Postgres, and transactions to manage application data.
We manually review bug reports and search for recent bugs marked as ``hard to reproduce,''
looking in particular for database-related bugs.

In MediaWiki, we observe that non-atomic update of multiple entries is a common cause of concurrency bugs
which \sn simplifies debugging.
For example, \emph{MW-44325}~\cite{wikimedia-44325} has a race condition where multiple concurrent edits of the same page can create duplicated site URL links,
which violates the uniqueness requirement of site URL links.
This bug is caused by non-atomic updates of the edited page object and the \texttt{SiteLink} table that stores page links,
but developers struggled to reproduce the bug and accurately locate the root cause.
The bug discussion included 33 developers, started in November 2012, closed and reopened multiple times, and was finally marked as resolved in October 2021.
Developers took such a long time to fix this bug because they mostly rely on production logs to figure out ways to reproduce it; however, their logs do not provide a full view of concurrent executions and data operations,
so developers have to guess the execution sequence that might lead to the error. 
By contrast, to debug this issue with \sn, developers can simply query \sn provenance data to inspect which requests edited the same page and created the duplicated URL links.
Then, developers can use \sn to replay the buggy requests to examine the interaction of relevant concurrent executions and their corresponding changes to the page object and the \texttt{SiteLink} table.

Another similar bug was
\emph{MW-39225}~\cite{wikimedia-39225} where the page edit handler rarely and randomly returns wrong article size changes in the presence of concurrent edits.
This bug is due to non-atomic page edits so the concurrent requests can see partially updated data.
Developers complained that the server logs were not helpful, and they had to guess ways to reproduce the bug, which led to a three-month-long discussion thread.
Similar to MW-44325, developers can use \sn to locate and reproduce this bug through declarative debugging and bug replay.

In Moodle, we found several more database-related time-of-check to time-to-use (TOCTOU) bugs similar to MDL-59854 (Section~\ref{sec:background}), which caused duplications in other database tables and can be debugged with \sn.
Sometimes, fixes to these bugs cause more bugs.
For example, the patch for MDL-59854 later caused another production error, MDL-60669~\cite{moodle-60669},
because it did not consider the corner case where duplications still exist in deleted courses, so restoring those courses raised errors.
\sn's retroactive programming feature can prevent these new bugs by helping developers validate bug fixes before applying them in production.
For example, to thoroughly test their bug fixes, developers can apply the modified code and serve past user requests directly related to this bug and other requests that may touch the same table.

\subsection{Security Bugs}
Prior work has shown that security bugs form a large and growing class of bugs in web applications~\cite{near2016finding}.
These bugs may allow attackers to illegally modify application state or exfiltrate sensitive user data from an application,
leading to severe security breaches.
However, investigating security issues is hard because developers need to trace across many request handlers running on different machines.
Currently, developers can only investigate these issues using system logs, but these logs may not provide sufficient information to track attacks~\cite{datta2022alastor}.
For example, logs usually contain error messages but do not provide visibility into how applications interact with databases.
By contrast, developers can easily query \sn's provenance database to check for access control violations or potential data exfiltrations.

\sn makes it easy to check web applications for violations of common access control patterns, as identified by Near and Jackson~\cite{near2016finding},
such as the Authentication pattern (only allowing logged-in users to read certain objects) and the User Profiles pattern (only users themselves can update their profiles).
We now demonstrate how \sn can help detect potential violations of the User Profiles pattern;
\sn can detect violations of the other patterns through similar means.
Specifically, we want to find all requests that illegally updated a profile (i.e., the request was not made by the profile owner).  We can do this with the following SQL query:
\begin{tcolorbox}[left=0pt,right=0pt,top=0pt,bottom=0pt,before skip=5pt,after skip=5pt]
	\begin{Verbatim}[fontfamily=lmtt,fontsize=\small]
SELECT Timestamp, ReqId, HandlerName
FROM Executions as E, ProfileEvents as P
  ON E.TxnId = P.TxnId
WHERE P.UserName != P.UpdatedBy AND P.Type = 'Update'
	\end{Verbatim}
\end{tcolorbox}

\sn can also support more complex forensics investigation queries such as detecting data exfiltration through workflows~\cite{datta2022alastor},
where attackers can leverage RPCs between handlers to move stolen data laterally through workflow executions and finally exfiltrate data over a seemingly valid workflow.
Since \sn traces the entire workflow of handler invocations that serve each request,
developers can query \sn provenance data to track all subsequent changes made by a request that improperly accessed sensitive data,
and determine if the data is exfiltrated.
\section{Challenges and Research Directions}
\label{sec:future}

\noindent\textbf{Debugging Performance and Data Issues.}
In this paper, we mainly focus on functionality bugs, but we hope to extend \sn to other categories such as
1) performance bugs where executions are correct but slow,
and 2) data quality bugs which lead to well-formed but incorrect data, caused mainly by human errors.
We can extend \sn to support more monitoring features like performance profiling to help developers troubleshoot performance bugs.
For example, commercial APM (application performance monitoring) tools such as Retrace~\cite{retrace} and New Relic~\cite{newrelic} allow developers to annotate their applications to obtain transaction traces for debugging slow queries. \sn can similarly augment its execution tracing to  record performance metrics such as latencies of individual handlers and end-to-end executions,
and store this information in a structured and queryable format.
We believe these automatically generated traces would make it far easier for developers to use data-driven performance debugging tools like Seer~\cite{gan2019seer}, which currently require extensive and tedious manual annotations.

We also believe \sn can simplify debugging data quality issues, because it captures all application data changes and accesses.
\sn can leverage prior work on data debugging.
For example, we may support data quality tests over \sn's provenance database to discover erroneous edits~\cite{mucslu2013data},
and find requests that caused data quality degradation.
One challenge is to balance the granularity of provenance recording and the accuracy of data quality testing.
We expect to extend \sn's bug replay to reconstruct more detailed provenance data retroactively~\cite{arab2014generic} without harming performance.

\vspace{0.5em}
\noindent\textbf{Handling Multiple Data Stores.}
Modern web applications and microservices may use multiple data stores for heterogeneous data.
For example, they can use a combination of a relational DBMS (e.g., Postgres, MySQL), a key-value store (e.g., Redis), a document store (e.g., MongoDB), and a search engine (e.g., ElasticSearch)~\cite{laigner2021data}.
It is challenging for these applications to use \sn because some data stores do not support transactions, and transaction logs of different stores are usually not aligned.
However, recent work has proposed transaction managers that support transactions across heterogeneous, even non-transactional, data stores~\cite{cherrygarcia,faria2021towards}.
Such transaction managers can also provide aligned transaction logs.
Therefore, we believe \sn can work for applications using multiple data stores if they adopt such cross-data store transaction managers. 

\vspace{0.5em}
\noindent\textbf{Guaranteeing Security and Privacy.}
Web applications often need to comply with regulations such as GDPR and CCPA to guarantee the security and privacy of sensitive user data.
This poses a challenge for TROD because TROD may log sensitive user data such as personally identifiable information (PII).
In order to effectively debug applications while respecting user privacy, \sn needs to let users completely remove any provenance data entry that potentially contains their personal information and support debugging from partial data.
Therefore, we plan to research ways to maintain non-sensitive but critical metadata and partially log encrypted data for debugging without violating regulations.

\section{Related Work}

\noindent\textbf{Provenance for Debugging.}
Prior work~\cite{provenancesurvey} has shown that provenance information can improve the reproducibility of database transactions and process debugging.
Similar to \sn, VisTrails Total Recall~\cite{chirigati2012towards} enables reproducibility for scientific workflows that interact with databases by combining workflow and data provenance.
GProM~\cite{arab2014generic} introduces a provenance middleware that interposes between user queries and databases;
it also relies on transaction logs and database time travel to reconstruct provenance on demand to reduce overhead.
GProM works for lower isolation levels, such as snapshot isolation and read-committed.
Watermelon~\cite{whittaker2018debugging} generalizes why-provenance to debugging distributed systems and proposes wat-provenance (why-across-time provenance), which also requires determinism.
Chen et al.~\cite{chen2017data} propose network provenance to debug computer networks at Internet scale.
However, prior work focuses either on database queries or application logic, and thus cannot faithfully replay database-backed applications or support retroactive modifications.
We believe that \sn could adapt and extend these ideas from prior provenance research to further optimize the tracing and representation of essential information for debugging distributed applications.

\vspace{0.5em}
\noindent\textbf{Bug Replay and Retroactive Programming.}
Arnold~\cite{devecsery2014eidetic} and OmniTable~\cite{quinn2022omnitable} capture detailed lineage of the entire system state at the granularity of individual instructions and every memory and register state of processes,
which allows them to reproduce any transient state in the past.
They do not support distributed applications, but \sn can borrow ideas from these systems to handle external calls and non-deterministic events.
Pensieve~\cite{zhang2017pensieve} reproduces distributed systems failures using system logs and application bytecode,
but it requires developers to provide sufficient information in the error logs.
\sn's retroactive programming is inspired by Retro-$\lambda$~\cite{meissner2018retro}, an event-sourced platform that supports retroactive programming for serverless applications,
but it only considers a single isolated microservice and does not support transactions. 
\section{Conclusion}
\label{sec:conclusion}
Debugging distributed online web applications is challenging.
Current practices require intensive manual logging and guesswork.
In this paper, we proposed \sn, a novel transaction-oriented debugger for distributed web applications.
We showed that \sn provides always-on tracing and declarative debugging, faithfully reproduces any concurrency bugs, and allows developers to test modified code retroactively on any past event.
We encouraged the database and systems communities to drastically rethink the synergy between the way people develop and debug applications.

\section*{Acknowledgements}
We thank the anonymous reviewers for their helpful feedback. We also thank members of the DBOS project for their insightful discussions to improve this work.
This work was supported by the Stanford Platform Lab and its industrial affiliates.

\bibliographystyle{ACM-Reference-Format}
\bibliography{references}


\begin{thebibliography}{30}


\ifx \showCODEN    \undefined \def \showCODEN     #1{\unskip}     \fi
\ifx \showDOI      \undefined \def \showDOI       #1{#1}\fi
\ifx \showISBNx    \undefined \def \showISBNx     #1{\unskip}     \fi
\ifx \showISBNxiii \undefined \def \showISBNxiii  #1{\unskip}     \fi
\ifx \showISSN     \undefined \def \showISSN      #1{\unskip}     \fi
\ifx \showLCCN     \undefined \def \showLCCN      #1{\unskip}     \fi
\ifx \shownote     \undefined \def \shownote      #1{#1}          \fi
\ifx \showarticletitle \undefined \def \showarticletitle #1{#1}   \fi
\ifx \showURL      \undefined \def \showURL       {\relax}        \fi
\providecommand\bibfield[2]{#2}
\providecommand\bibinfo[2]{#2}
\providecommand\natexlab[1]{#1}
\providecommand\showeprint[2][]{arXiv:#2}

\bibitem[Arab et~al\mbox{.}(2014)]%
        {arab2014generic}
\bibfield{author}{\bibinfo{person}{Bahareh Arab}, \bibinfo{person}{Dieter
  Gawlick}, \bibinfo{person}{Venkatesh Radhakrishnan}, \bibinfo{person}{Hao
  Guo}, {and} \bibinfo{person}{Boris Glavic}.} \bibinfo{year}{2014}\natexlab{}.
\newblock \showarticletitle{{A Generic Provenance Middleware for Queries,
  Updates, and Transactions}}. In \bibinfo{booktitle}{\emph{TaPP 2014}}.
\newblock


\bibitem[AWS(2022)]%
        {awslambda}
\bibfield{author}{\bibinfo{person}{AWS}.} \bibinfo{year}{2022}\natexlab{}.
\newblock \bibinfo{title}{{AWS Lambda}}.
\newblock
\newblock
\newblock
\shownote{\url{https://aws.amazon.com/lambda/}}.


\bibitem[Chen et~al\mbox{.}(2017)]%
        {chen2017data}
\bibfield{author}{\bibinfo{person}{Ang Chen}, \bibinfo{person}{Yang Wu},
  \bibinfo{person}{Andreas Haeberlen}, \bibinfo{person}{Boon~Thau Loo}, {and}
  \bibinfo{person}{Wenchao Zhou}.} \bibinfo{year}{2017}\natexlab{}.
\newblock \showarticletitle{{Data Provenance at Internet Scale: Architecture,
  Experiences, and the Road Ahead}}. In \bibinfo{booktitle}{\emph{CIDR 2017}}.
\newblock


\bibitem[Chirigati and Freire(2012)]%
        {chirigati2012towards}
\bibfield{author}{\bibinfo{person}{Fernando Chirigati} {and}
  \bibinfo{person}{Juliana Freire}.} \bibinfo{year}{2012}\natexlab{}.
\newblock \showarticletitle{{Towards Integrating Workflow and Database
  Provenance}}. In \bibinfo{booktitle}{\emph{IPAW 2012}}. Springer,
  \bibinfo{pages}{11--23}.
\newblock


\bibitem[Datta et~al\mbox{.}(2022)]%
        {datta2022alastor}
\bibfield{author}{\bibinfo{person}{Pubali Datta}, \bibinfo{person}{Isaac
  Polinsky}, \bibinfo{person}{Muhammad~Adil Inam}, \bibinfo{person}{Adam
  Bates}, {and} \bibinfo{person}{William Enck}.}
  \bibinfo{year}{2022}\natexlab{}.
\newblock \showarticletitle{{ALASTOR: Reconstructing the Provenance of
  Serverless Intrusions}}. In \bibinfo{booktitle}{\emph{USENIX Security 22}}.
  \bibinfo{pages}{2443--2460}.
\newblock


\bibitem[Devecsery et~al\mbox{.}(2014)]%
        {devecsery2014eidetic}
\bibfield{author}{\bibinfo{person}{David Devecsery}, \bibinfo{person}{Michael
  Chow}, \bibinfo{person}{Xianzheng Dou}, \bibinfo{person}{Jason Flinn}, {and}
  \bibinfo{person}{Peter~M. Chen}.} \bibinfo{year}{2014}\natexlab{}.
\newblock \showarticletitle{{Eidetic Systems}}. In
  \bibinfo{booktitle}{\emph{OSDI 2014}}. \bibinfo{address}{Broomfield, CO},
  \bibinfo{pages}{525--540}.
\newblock
\showISBNx{978-1-931971-16-4}


\bibitem[Dey et~al\mbox{.}(2015)]%
        {cherrygarcia}
\bibfield{author}{\bibinfo{person}{Akon Dey}, \bibinfo{person}{Alan Fekete},
  {and} \bibinfo{person}{Uwe R{\"o}hm}.} \bibinfo{year}{2015}\natexlab{}.
\newblock \showarticletitle{Scalable distributed transactions across
  heterogeneous stores}. In \bibinfo{booktitle}{\emph{ICDE 2015}}.
  \bibinfo{pages}{125--136}.
\newblock


\bibitem[Erfani~Joorabchi et~al\mbox{.}(2014)]%
        {erfani2014works}
\bibfield{author}{\bibinfo{person}{Mona Erfani~Joorabchi},
  \bibinfo{person}{Mehdi Mirzaaghaei}, {and} \bibinfo{person}{Ali Mesbah}.}
  \bibinfo{year}{2014}\natexlab{}.
\newblock \showarticletitle{Works for me! characterizing non-reproducible bug
  reports}. In \bibinfo{booktitle}{\emph{MSR 2014}}. \bibinfo{pages}{62--71}.
\newblock


\bibitem[Faria et~al\mbox{.}(2021)]%
        {faria2021towards}
\bibfield{author}{\bibinfo{person}{Nuno Faria}, \bibinfo{person}{Jos{\'e}
  Pereira}, \bibinfo{person}{Ana~Nunes Alonso}, {and} \bibinfo{person}{Ricardo
  Vila{\c{c}}a}.} \bibinfo{year}{2021}\natexlab{}.
\newblock \showarticletitle{Towards generic fine-grained transaction isolation
  in polystores}.
\newblock In \bibinfo{booktitle}{\emph{Poly 2021}}. \bibinfo{pages}{29--42}.
\newblock


\bibitem[Gan et~al\mbox{.}(2019)]%
        {gan2019seer}
\bibfield{author}{\bibinfo{person}{Yu Gan}, \bibinfo{person}{Yanqi Zhang},
  \bibinfo{person}{Kelvin Hu}, \bibinfo{person}{Dailun Cheng},
  \bibinfo{person}{Yuan He}, \bibinfo{person}{Meghna Pancholi}, {and}
  \bibinfo{person}{Christina Delimitrou}.} \bibinfo{year}{2019}\natexlab{}.
\newblock \showarticletitle{{Seer: Leveraging big data to navigate the
  complexity of performance debugging in cloud microservices}}. In
  \bibinfo{booktitle}{\emph{ASPLOS 2019}}. \bibinfo{pages}{19--33}.
\newblock


\bibitem[Gray(1986)]%
        {gray1986computers}
\bibfield{author}{\bibinfo{person}{Jim Gray}.} \bibinfo{year}{1986}\natexlab{}.
\newblock \showarticletitle{Why do computers stop and what can be done about
  it?}. In \bibinfo{booktitle}{\emph{SRDS 1986}}. Los Angeles, CA, USA,
  \bibinfo{pages}{3--12}.
\newblock


\bibitem[Herlihy and Wing(1990)]%
        {linearizability}
\bibfield{author}{\bibinfo{person}{Maurice~P. Herlihy} {and}
  \bibinfo{person}{Jeannette~M. Wing}.} \bibinfo{year}{1990}\natexlab{}.
\newblock \showarticletitle{Linearizability: A Correctness Condition for
  Concurrent Objects}.
\newblock \bibinfo{journal}{\emph{ACM Trans. Program. Lang. Syst.}}
  \bibinfo{volume}{12}, \bibinfo{number}{3} (\bibinfo{date}{jul}
  \bibinfo{year}{1990}), \bibinfo{pages}{463–492}.
\newblock
\showISSN{0164-0925}
\urldef\tempurl%
\url{https://doi.org/10.1145/78969.78972}
\showDOI{\tempurl}


\bibitem[Herschel et~al\mbox{.}(2017)]%
        {provenancesurvey}
\bibfield{author}{\bibinfo{person}{Melanie Herschel}, \bibinfo{person}{Ralf
  Diestelk{\"a}mper}, {and} \bibinfo{person}{Houssem~Ben Lahmar}.}
  \bibinfo{year}{2017}\natexlab{}.
\newblock \showarticletitle{A survey on provenance: What for? What form? What
  from?}
\newblock \bibinfo{journal}{\emph{The VLDB Journal}} \bibinfo{volume}{26},
  \bibinfo{number}{6} (\bibinfo{year}{2017}), \bibinfo{pages}{881--906}.
\newblock


\bibitem[Kraft* et~al\mbox{.}(2022)]%
        {apiary}
\bibfield{author}{\bibinfo{person}{Peter Kraft*}, \bibinfo{person}{Qian Li*},
  \bibinfo{person}{Kostis Kaffes}, \bibinfo{person}{Athinagoras Skiadopoulos},
  \bibinfo{person}{Deeptaanshu Kumar}, \bibinfo{person}{Danny Cho},
  \bibinfo{person}{Jason Li}, \bibinfo{person}{Robert Redmond},
  \bibinfo{person}{Nathan Weckwerth}, \bibinfo{person}{Brian Xia},
  \bibinfo{person}{Peter Bailis}, \bibinfo{person}{Michael Cafarella},
  \bibinfo{person}{Goetz Graefe}, \bibinfo{person}{Jeremy Kepner},
  \bibinfo{person}{Christos Kozyrakis}, \bibinfo{person}{Michael Stonebraker},
  \bibinfo{person}{Lalith Suresh}, \bibinfo{person}{Xiangyao Yu}, {and}
  \bibinfo{person}{Matei Zaharia}.} \bibinfo{year}{2022}\natexlab{}.
\newblock \bibinfo{title}{Apiary: A DBMS-Backed Transactional
  Function-as-a-Service Framework}.
\newblock
\newblock
\urldef\tempurl%
\url{https://doi.org/10.48550/ARXIV.2208.13068}
\showDOI{\tempurl}


\bibitem[Laigner et~al\mbox{.}(2021)]%
        {laigner2021data}
\bibfield{author}{\bibinfo{person}{Rodrigo Laigner}, \bibinfo{person}{Yongluan
  Zhou}, \bibinfo{person}{Marcos Antonio~Vaz Salles}, \bibinfo{person}{Yijian
  Liu}, {and} \bibinfo{person}{Marcos Kalinowski}.}
  \bibinfo{year}{2021}\natexlab{}.
\newblock \showarticletitle{{Data Management in Microservices: State of the
  Practice, Challenges, and Research Directions}}.
\newblock \bibinfo{journal}{\emph{Proc. VLDB Endow.}} \bibinfo{volume}{14},
  \bibinfo{number}{13} (\bibinfo{year}{2021}), \bibinfo{pages}{3348–3361}.
\newblock
\showISSN{2150-8097}


\bibitem[Li et~al\mbox{.}(2022)]%
        {2022dbos-progress}
\bibfield{author}{\bibinfo{person}{Qian Li}, \bibinfo{person}{Peter Kraft},
  \bibinfo{person}{Kostis Kaffes}, \bibinfo{person}{Athinagoras Skiadopoulos},
  \bibinfo{person}{Deeptaanshu Kumar}, \bibinfo{person}{Jason Li},
  \bibinfo{person}{Michael Cafarella}, \bibinfo{person}{Goetz Graefe},
  \bibinfo{person}{Jeremy Kepner}, \bibinfo{person}{Christos Kozyrakis},
  \bibinfo{person}{Michael Stonebraker}, \bibinfo{person}{Lalith Suresh}, {and}
  \bibinfo{person}{Matei Zaharia}.} \bibinfo{year}{2022}\natexlab{}.
\newblock \showarticletitle{{A Progress Report on DBOS: A Database-oriented
  Operating System}}. In \bibinfo{booktitle}{\emph{CIDR 2022}}.
\newblock


\bibitem[MediaWiki(2012)]%
        {wikimedia-44325}
\bibfield{author}{\bibinfo{person}{MediaWiki}.}
  \bibinfo{year}{2012}\natexlab{}.
\newblock \bibinfo{title}{{Prevent creation of items having the same sitelinks
  (duplicates)}}.
\newblock
\newblock
\newblock
\shownote{\url{https://phabricator.wikimedia.org/T44325}}.


\bibitem[MediaWiki(2014)]%
        {wikimedia-39225}
\bibfield{author}{\bibinfo{person}{MediaWiki}.}
  \bibinfo{year}{2014}\natexlab{}.
\newblock \bibinfo{title}{{Several history entries for the same content and
  watchlists showing wrong article size changes}}.
\newblock
\newblock
\newblock
\shownote{\url{https://phabricator.wikimedia.org/T39225}}.


\bibitem[Meissner et~al\mbox{.}(2018)]%
        {meissner2018retro}
\bibfield{author}{\bibinfo{person}{Dominik Meissner}, \bibinfo{person}{Benjamin
  Erb}, \bibinfo{person}{Frank Kargl}, {and} \bibinfo{person}{Matthias Tichy}.}
  \bibinfo{year}{2018}\natexlab{}.
\newblock \showarticletitle{{Retro-$\lambda$: An event-sourced platform for
  serverless applications with retroactive computing support}}. In
  \bibinfo{booktitle}{\emph{DEBS 2018}}. \bibinfo{pages}{76--87}.
\newblock


\bibitem[Moodle(2017a)]%
        {moodle-60669}
\bibfield{author}{\bibinfo{person}{Moodle}.} \bibinfo{year}{2017}\natexlab{a}.
\newblock \bibinfo{title}{{Course restore can fail if it includes discussion
  forums}}.
\newblock
\newblock
\newblock
\shownote{\url{https://tracker.moodle.org/browse/MDL-60669}}.


\bibitem[Moodle(2017b)]%
        {moodle-59854}
\bibfield{author}{\bibinfo{person}{Moodle}.} \bibinfo{year}{2017}\natexlab{b}.
\newblock \bibinfo{title}{{Duplicate forum subscriptions due to race
  conditions}}.
\newblock
\newblock
\newblock
\shownote{\url{https://tracker.moodle.org/browse/MDL-59854}}.


\bibitem[Mu{\c{s}}lu et~al\mbox{.}(2013)]%
        {mucslu2013data}
\bibfield{author}{\bibinfo{person}{K{\i}van{\c{c}} Mu{\c{s}}lu},
  \bibinfo{person}{Yuriy Brun}, {and} \bibinfo{person}{Alexandra Meliou}.}
  \bibinfo{year}{2013}\natexlab{}.
\newblock \showarticletitle{Data debugging with continuous testing}. In
  \bibinfo{booktitle}{\emph{ESEC/FSE 2013}}. \bibinfo{pages}{631--634}.
\newblock


\bibitem[Near and Jackson(2016)]%
        {near2016finding}
\bibfield{author}{\bibinfo{person}{Joseph~P Near} {and} \bibinfo{person}{Daniel
  Jackson}.} \bibinfo{year}{2016}\natexlab{}.
\newblock \showarticletitle{Finding security bugs in web applications using a
  catalog of access control patterns}. In \bibinfo{booktitle}{\emph{ICSE
  2016}}. \bibinfo{pages}{947--958}.
\newblock


\bibitem[Qiu et~al\mbox{.}(2022)]%
        {qiu2022deep}
\bibfield{author}{\bibinfo{person}{Zhengyi Qiu}, \bibinfo{person}{Shudi Shao},
  \bibinfo{person}{Qi Zhao}, \bibinfo{person}{Hassan~Ali Khan},
  \bibinfo{person}{Xinning Hui}, {and} \bibinfo{person}{Guoliang Jin}.}
  \bibinfo{year}{2022}\natexlab{}.
\newblock \showarticletitle{{A Deep Study of the Effects and Fixes of
  Server-Side Request Races in Web Applications}}. In
  \bibinfo{booktitle}{\emph{MSR 2022}}. \bibinfo{pages}{744--756}.
\newblock


\bibitem[Quinn et~al\mbox{.}(2022)]%
        {quinn2022omnitable}
\bibfield{author}{\bibinfo{person}{Andrew Quinn}, \bibinfo{person}{Jason
  Flinn}, \bibinfo{person}{Michael Cafarella}, {and} \bibinfo{person}{Baris
  Kasikci}.} \bibinfo{year}{2022}\natexlab{}.
\newblock \showarticletitle{Debugging the {OmniTable} Way}. In
  \bibinfo{booktitle}{\emph{OSDI 2022}}. \bibinfo{address}{Carlsbad, CA},
  \bibinfo{pages}{357--373}.
\newblock
\showISBNx{978-1-939133-28-1}


\bibitem[Relic(2022)]%
        {newrelic}
\bibfield{author}{\bibinfo{person}{New Relic}.}
  \bibinfo{year}{2022}\natexlab{}.
\newblock \bibinfo{title}{{Transactions in New Relic's APM}}.
\newblock
\newblock
\newblock
\shownote{\url{https://docs.newrelic.com/docs/apm/transactions/intro-transactions/transactions-new-relic-apm/}}.


\bibitem[Stackify(2022)]%
        {retrace}
\bibfield{author}{\bibinfo{person}{Stackify}.} \bibinfo{year}{2022}\natexlab{}.
\newblock \bibinfo{title}{{Centralized Log Management}}.
\newblock
\newblock
\newblock
\shownote{\url{https://stackify.com/retrace-log-management/}}.


\bibitem[Whittaker et~al\mbox{.}(2018)]%
        {whittaker2018debugging}
\bibfield{author}{\bibinfo{person}{Michael Whittaker},
  \bibinfo{person}{Cristina Teodoropol}, \bibinfo{person}{Peter Alvaro}, {and}
  \bibinfo{person}{Joseph~M Hellerstein}.} \bibinfo{year}{2018}\natexlab{}.
\newblock \showarticletitle{Debugging distributed systems with why-across-time
  provenance}. In \bibinfo{booktitle}{\emph{SoCC 2018}}.
  \bibinfo{pages}{333--346}.
\newblock


\bibitem[Zhang et~al\mbox{.}(2017)]%
        {zhang2017pensieve}
\bibfield{author}{\bibinfo{person}{Yongle Zhang}, \bibinfo{person}{Serguei
  Makarov}, \bibinfo{person}{Xiang Ren}, \bibinfo{person}{David Lion}, {and}
  \bibinfo{person}{Ding Yuan}.} \bibinfo{year}{2017}\natexlab{}.
\newblock \showarticletitle{{Pensieve: Non-Intrusive Failure Reproduction for
  Distributed Systems Using the Event Chaining Approach}}. In
  \bibinfo{booktitle}{\emph{SOSP 2017}}. \bibinfo{pages}{19–33}.
\newblock
\showISBNx{9781450350853}


\bibitem[Zhou et~al\mbox{.}(2021)]%
        {zhou2021foundationdb}
\bibfield{author}{\bibinfo{person}{Jingyu Zhou}, \bibinfo{person}{Meng Xu},
  \bibinfo{person}{Alexander Shraer}, \bibinfo{person}{Bala Namasivayam},
  \bibinfo{person}{Alex Miller}, {et~al\mbox{.}}}
  \bibinfo{year}{2021}\natexlab{}.
\newblock \showarticletitle{{Foundationdb: A distributed unbundled
  transactional key value store}}. In \bibinfo{booktitle}{\emph{SIGMOD 2021}}.
  \bibinfo{pages}{2653--2666}.
\newblock


\end{thebibliography}

\appendix

\end{document}